\documentclass[letter,twocolumn]{jpsj3}
\usepackage{txfonts}
\usepackage{bm}

\title{A Monomer Mott Insulator (BEDT-TTF)Cu[N(CN)$_2$]$_2$ \\
as a Potential Nodal Line System}

\author{Naoki Yoneyama$^1$\thanks{nyoneyama@yamanashi.ac.jp}, Muhammad Khalish Nuryadin$^2$, Takao Tsumuraya$^3$, Satoshi Iguchi$^2$,\\
 Takahiro Takei$^4$, Nobuhiro Kumada$^4$, Masanori Nagao$^4$, Isao Tanaka$^4$, and Takahiko Sasaki$^2$}

\inst{$^1$Graduate Faculty of Interdisciplinary Research, University of Yamanashi, Kofu 400-8511, Japan \\
$^2$Institute for Materials Research, Tohoku University, Sendai 980-8577, Japan \\
$^3$Priority Organization for Innovation and Excellence, Kumamoto University, Kumamoto 860-8555, Japan \\
$^4$Center for Crystal Science and Technology, University of Yamanashi, Kofu 400-0021, Japan \\
} 

\abst{
We report the band structure calculations and the experimental results 
of resistivity and magnetic susceptibility in a spin-1/2 
(BEDT-TTF)$^{\bullet +}$ monomer Mott insulator (BEDT-TTF)Cu[N(CN)$_2$]$_2$. 
The band calculations indicate a Dirac semimetal state with nodal lines at the Fermi level.
The resistivity and the magnetic susceptibility as functions of temperature 
are well interpreted in terms of the monomer Mott insulating state
instead of the expected semimetal state probably owing to strong 
electron correlation.
In addition, we find that an Arrhenius-type steep reduction of the paramagnetic susceptibility
appears below approximately 25 K, which indicates a spin-singlet ground state.
}

\begin{document}
\maketitle

The electronic band structures with linear dispersion
have recently been interested because of the mass-less Dirac nature of 
the fermion carriers.
When the Fermi level is located near the center of the dispersion,
the Dirac semimetal (DS) state can be achieved.
The discovery of the quantum Hall effect in graphene\cite{Novoselov} as a 
two-dimensional (2D) DS with the cone-type dispersion has begun to 
attract many studies.
Especially, organic molecular based compounds are good platform to investigate the
DS state in the bulk system.
The first DS state in organics is recognized in a 2D layered 
system of $\alpha$-(BEDT-TTF)$_2$I$_3$\cite{Katayama}, and its relatives are recently developing\cite{STF,BETS},
where BEDT-TTF is bis(ethylenedithio)tetrathiafulvalene (abbreviated below as ET).
The electronic band near the Fermi level in these systems is 
derived from the HOMO orbital of the ET molecule.
On the other hand, in the single-component conductor [Pd(dddt)$_2$], 
the DS state with nodal lines near the Fermi level comes from the HOMO 
and LUMO multi bands, which arises under pressure.\cite{Pddddt2}
Moreover, [Pt(dmdt)$_2$] realizes the node line DS state at ambient pressure.\cite{Ptdmdt2}
Another interesting DS candidate is the three-dimensional (3D) diamond lattice
system of (ET)Ag$_4$(CN)$_5$ with a 1/2-filled band.\cite{AgCN1,AgCN2}

We here focus on an ET salt with Cu(I) dicyanamide counter anion,
(ET)Cu[N(CN)$_2$]$_2$,\cite{ETDCA1}
which will be classified as a modified analogue of (ET)Ag$_4$(CN)$_5$.
(ET)Cu[N(CN)$_2$]$_2$ is a by-product of the organic dimer Mott 
insulator/superconductor $\kappa$-(ET)$_2$Cu[N(CN)$_2$]Cl ($T_c=12.8$ K 
under 0.3 kbar)\cite{ETDCA1}, one of the most notable 2D-layered 
conducting systems composed of strongly dimerized ET molecules.
In contrast to the dimer Mott system, (ET)Cu[N(CN)$_2$]$_2$ could be 
considered as a spin-1/2 ET$^{\bullet +}$ monomer Mott insulator.
There is no 2D conducting sheet structure in (ET)Cu[N(CN)$_2$]$_2$, in
which a peculiar 3D anisotropic diamond-like network 
of the ET molecules has been overlooked for decades\cite{ETDCA2}.

In the present paper, we investigate the band structure at room 
temperature (RT), revealing that 
the uniform zigzag chain with dihedral inter-chain (anisotropic diamond-like) interaction
gives rise to a potential DS state with nodal lines.
By means of X-ray crystal structure analysis, dc resistivity, and static magnetic 
susceptibility measurements,
the paramagnetic insulating properties are interpreted in 
terms of the monomer Mott insulating state.

Single crystals of (ET)Cu[N(CN)$_2$]$_2$ were grown by conventional electrochemical method\cite{ETDCA1}.
The black needle-like crystalline shape with a typical dimension of 
$2\times0.05\times0.02$ mm$^3$ is easily distinguished by eyes from 
the thick platelet by-product of $\kappa$-(ET)$_2$Cu[N(CN)$_2$]Cl.
We performed X-ray structural analysis at 296 
and 100 K with different single crystals for temperature (Rigaku, XtaLAB mini and VariMax DW). 
For band calculations we adopt the crystallographic data taken from the literature,\cite{ETDCA1}
where the coordinates of H and N atoms were optimized for the DFT calculation.
In the tight-binding model, the intermolecular overlaps between the HOMO orbitals of ET were
calculated on the basis of the extended H\"{u}ckel method.\cite{BandCalc}
In the DFT band calculations, we employed projected augmented-wave pseudopotentials~\cite{PAW1994, DALCORSO_PP}
with plane wave basis sets implemented in~{\sc Quantum ESPRESSO}.~\cite{QE2017}
The cutoff energies for plane waves and charge densities were set at 45 and 488 Ry, respectively.
The exchange-correlation functional is the generalized gradient approximation by Perdew, Burke, and Ernzerhof.~\cite{GGA_PBE}
The dimensions of the $\bm{k}$-point mesh are $12\times12\times12$.
The dc resistivity measurements at ambient pressure were carried out by
means of conventional two- or four-terminal method,
while quasi four-terminal method\cite{q4terminal} with constant current
of 0.1 $\mu$A was 
adopted under pressure.
A CuBe clamp-type hydrostatic pressure cell was used for 
the sample in the $E \parallel c$ configuration up to 1.8 GPa with 
Daphne 7373 oil. 
The magnitude of pressure labeled below was estimated from the load 
gauge of the press machine at RT.
A calibration of pressure performed distinctively using the superconducting transition of Pb
showed the pressure loosen of approximately 0.3 GPa (at $\sim$ 10 K)
until 2 GPa.
Temperature was controlled with a PPMS system (Quantum Design, Dynacool).
The temperature dependence of the static magnetic susceptibility 
was measured by a SQUID magnetometer (Quantum 
Design, MPMS-XL) using single crystals with a total of 1.1 mg. 
A batch of samples was placed inside a poly-acetal rod as the sample holder. 
The samples were inserted into the center of the rod without using any grease
 and the long axis of multiple crystals was positioned to be
 perpendicular to the applied field ($H \perp c$).
We obtained the magnetic susceptibility ($\chi_\textrm{dc}$)
after subtraction of the diamagnetic contribution using the Pascal's constant, $\chi_\textrm{dia}=-3.6 
\times 10^{-4}$ emu/mol, from the measured data.

First it will be useful to describe the crystal structure.
The monoclinic ($C2/c$) crystal structure reported previously\cite{ETDCA1} is well 
reproduced at 296 K, and no significant variation is found at 100 K.\cite{crystaldata}
One-half crystallographically independent ET molecule exists in the 
crystal structure of (ET)Cu[N(CN)$_2$]$_2$.
The formal charge of ET is +1 because of the total valence of the anion 
(Cu(I)[N(CN)$_2$]$_2$)$^{-1}$, 
and is in good accordance with the charge estimated from the intra-molecular bond length of ET
at both 100 and 296 K.
Figure 1(a) shows the crystal structure viewed from the ET
molecular-long axis.
The ET molecules form a 1D zigzag chain at regular 
intervals along the $c$-axis with the nearest intermolecular (diagonally 
side-by-side) S-S contact (bold red line, $t$). 
There is a two-fold rotation axis perpendicular to the ET molecular plane,
which guarantees the 1D chain to be uniform.
In addition to the intra-chain coupling, the second-nearest interaction (blue wedge symbol, $t'$)
connects an ET with two others on the neighboring chains.
Namely all the ET molecules are in distorted tetrahedral coordination geometry.
The schematic view of the 3D network is shown in Fig. 1(b).
Gray spheres (centroids of ET) construct the uniform 1D zigzag (red) chain along the $c$-axis,
which is surrounded by four other chains via the (blue) couplings;
this corresponds to the strongly anisotropic (almost 1D) diamond-like structure.

\begin{figure}[b]
\begin{center}
\includegraphics[width=6.5cm,pagebox=cropbox,clip]{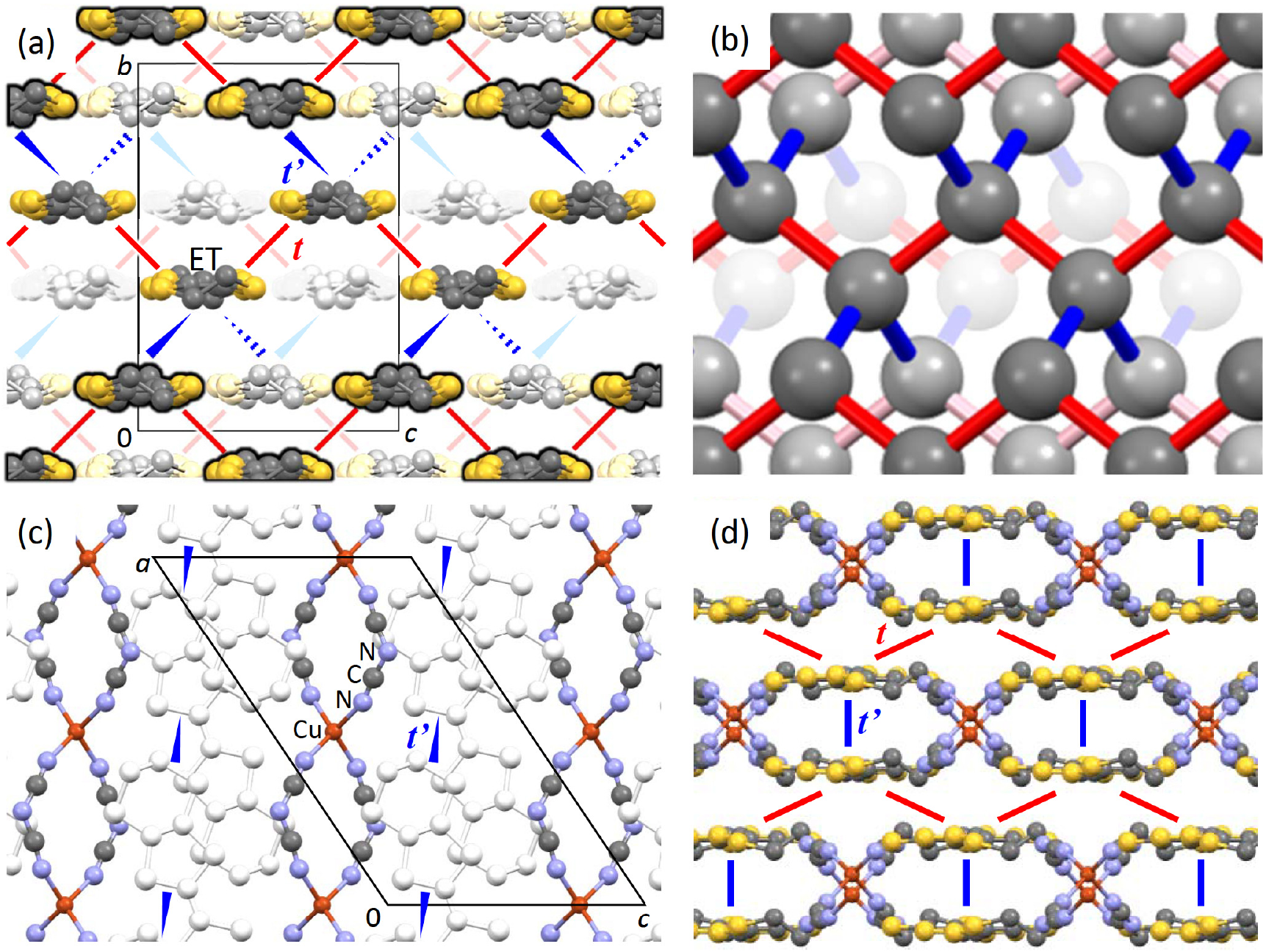}
 \caption{\label{Fig:structure} (Color online) Crystal structure of (BEDT-TTF)Cu[N(CN)$_2$]$_2$.
(a) View from the ET molecular long-axis shows the uniform 1D zigzag chains along the $c$-axis.
The colored 1D chain through the middle of the unit cell is located on the $bc$ plane.
The bold and pale-colored ET motifs are in the front and rear sides, respectively.
Anions are omitted for simplicity.
(b) A diamond-like lattice corresponding to the ET stacking shown in (a).
(c) View from the $b$-axis, showing the polymeric anion chain along the $a+c$ direction in color.
(d) View along the anion chain shown in (c).
}
\end{center}
\end{figure}
The polymeric anion is composed of a tetrahedral coordinate 
Cu(I) atom bridged by two bent dicyanamide anions (Fig. 1(c)) along the $a+c$ direction.
Because of the closed shell structure expected in the anion, it is 
almost likely that the anion does not contribute to the electronic state 
near the Fermi level and the properties such as conductivity and paramagnetism.
Figure 1(d), the view from the $a+c$ direction, indicates that 
the diamond-like packing of ET contains the columnar cavity occupied by the polymeric anions,
resulting in the unusual donor/anion mixed stacking structure. 

We next explain the band structure calculated using the tight-binding model.
The transfer integrals $t$ and $t'$ 
are estimated to be $-0.192$ and $+0.0229$ eV, respectively.
Almost the same amounts of $t (=-0.193$ eV) and $t' (=+0.0227$ eV) using our
X-ray data at 100 K are obtained, indicating that
the band picture presented below is safely preserved at least down to 100 K.
The conventional (monoclinic C) unit cell is reduced to the primitive cell 
with the lattice vectors $\bm{a}'=(\bm{a}-\bm{b})/2$, 
$\bm{b}'=(\bm{a}+\bm{b})/2$, and $\bm{c}'=\bm{c}$, as depicted in Fig. 2(a).
The primitive cell constants are $a'=b'=10.886$ {\AA}, 
$\alpha'=\beta'=115.45$\textdegree, and $\gamma'=79.63$\textdegree
($V'=V/2=993.2$ {\AA}$^3$ and $Z'=Z/2=2$).
The corresponding first Brillouin zone (BZ) is shown in Fig. 2(b) and
we use the primitive cell for the band calculations.

\begin{figure}[b]
\begin{center}
\includegraphics[width=6.5cm,pagebox=cropbox,clip]{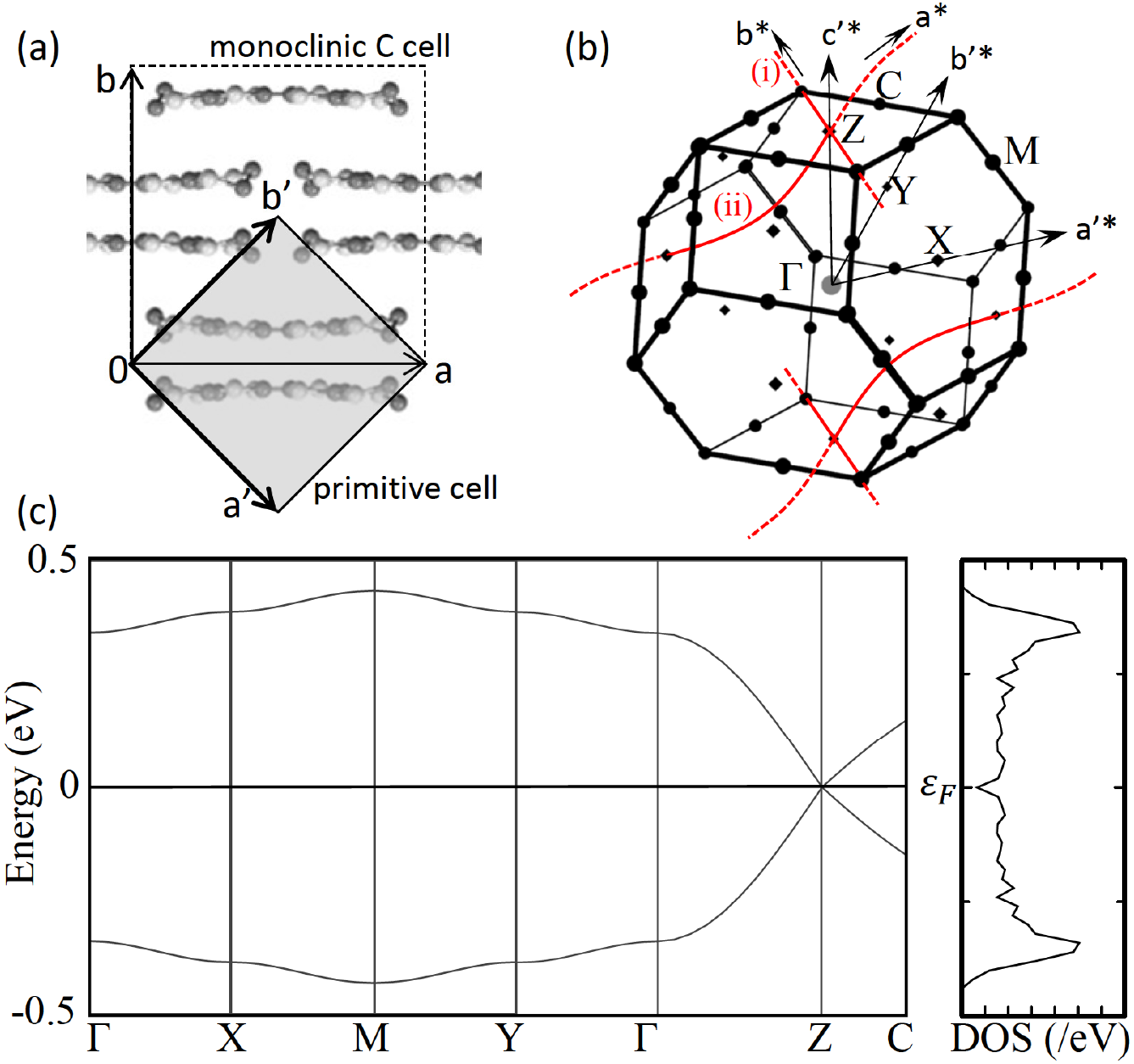}
 \caption{\label{Fig:band1} (Color online) (a) Primitive unit cell ($a'$, $b'$) projected on the 
 (001) plane of the conventional monoclinic C cell.
 (b) First Brillouin zone for the primitive cell with two nodal lines (i) 
 and (ii). The solid and broken red curves express inside and 
 outside of the BZ, respectively.
 (c) Band structure and the density of states (DOS) calculated within 
 the tight-binding approximation.
}
\end{center}
\end{figure}
On the basis of the standard tight-binding approximation,
the matrix elements of the $2\times2$ secular equation are $H_\textrm{AA}=H_\textrm{BB}=0$ and $H_\textrm{AB}=H^*_\textrm{BA}=t+te^{-ik_{c'}c'}+t'e^{ik_{a'}a'}+t'e^{-ik_{b'}b'-ik_{c'}c'}$.
The energy dispersion is then obtained as
\begin{equation}
E_{\pm}(\bm{k})=
\pm2\sqrt{(t\alpha_k+t'\beta_k)^2+2tt'\alpha_k\beta_k[\cos(a'k_{a'}-b'k_{b'})-1]},
\end{equation}
where $\alpha_k=\cos(k_{c'}c'/2)$ and $\beta_k=\cos(k_{a'}a'/2+k_{b'}b'/2+k_{c'}c'/2)$.
There are two band dispersions corresponding to the two equivalent ET molecules.
Taking account of one electron-transfer from the HOMO of ET to the charge 
compensating anion, the Fermi level is located at the middle of the HOMO band,
resulting in the 1/2-filled band.
The empty upper ($E_+$) and fully filled lower ($E_-$) bands are symmetric with respect to 
the Fermi level and they contact at $Z$ $(k_{a'},k_{b'},k_{c'})=(0,0,\pi/c')$ as depicted in Fig. 2(c). 
This band degeneracy at $Z$ comes from the zigzag uniform chain structure; 
by neglecting the small $t' (\ll t)$, 
one can reduce Eq. (1) to $E_{\pm}(\bm{k}) \approx \pm 2|t\cos(k_{c'}c'/2)|$,
leading to 1D ``nodal'' Fermi surfaces $(0 0 (\pm\pi/c'))$ on $Z$
with linear energy dispersion at around the 1st BZ.
In the present system, the weak $t'$ breaks the 1D Fermi degeneracy and
contributes to leave nodal lines orthogonally crossing at $Z$ as 
described below.
The two nodal lines (i) and (ii) are obtained from Eq. (1) by using the 
condition $E_{\pm}(\bm{k})=0$\cite{comment_on_node}: (i) is $k_{c'}c'=\pi$ and $k_{a'}a'+k_{b'}b'=0$, and 
(ii) $t\cos(k_{c'}c'/2)+t'\cos(k_{a'}a'+k_{c'}c'/2)=0$ and $k_{a'}a'-k_{b'}b'=0$.
The nodal lines are schematically depicted in Fig. 2(b).
The node (i) is linear along the $b^* (=-a'^*+b'^*)$-axis on the 1st BZ, while 
the node (ii) is bending along the $a^*(=a'^*+b'^*)$ 
direction on the $(1\bar{1}0)$ plane, that is the $c^*a^*$-plane in the monoclinic cell. 
As a result, strictly speaking, the present nodal feature originates not from the diamond-like
structure ($t'$) but from the zigzag uniform stacking along the $c$-axis 
($t$), corresponding to the existence of two equivalent sites in a 
primitive unit cell.
Thus, if there is a slight modification in the uniform zigzag chain 
such as a perturbative dimerization of ET molecules,
gap formation on all the nodal lines should easily occur.

Our tight-binding model based on the extended H\"{u}ckel method generally reproduces 
the DFT band structure shown in Fig. 3.
Although the bands derived from the anion orbitals overlap the lower HOMO band,
they does not contribute to the formation of the Dirac-type dispersion.

\begin{figure}[b]
\begin{center}
\includegraphics[width=6.5cm,pagebox=cropbox,clip]{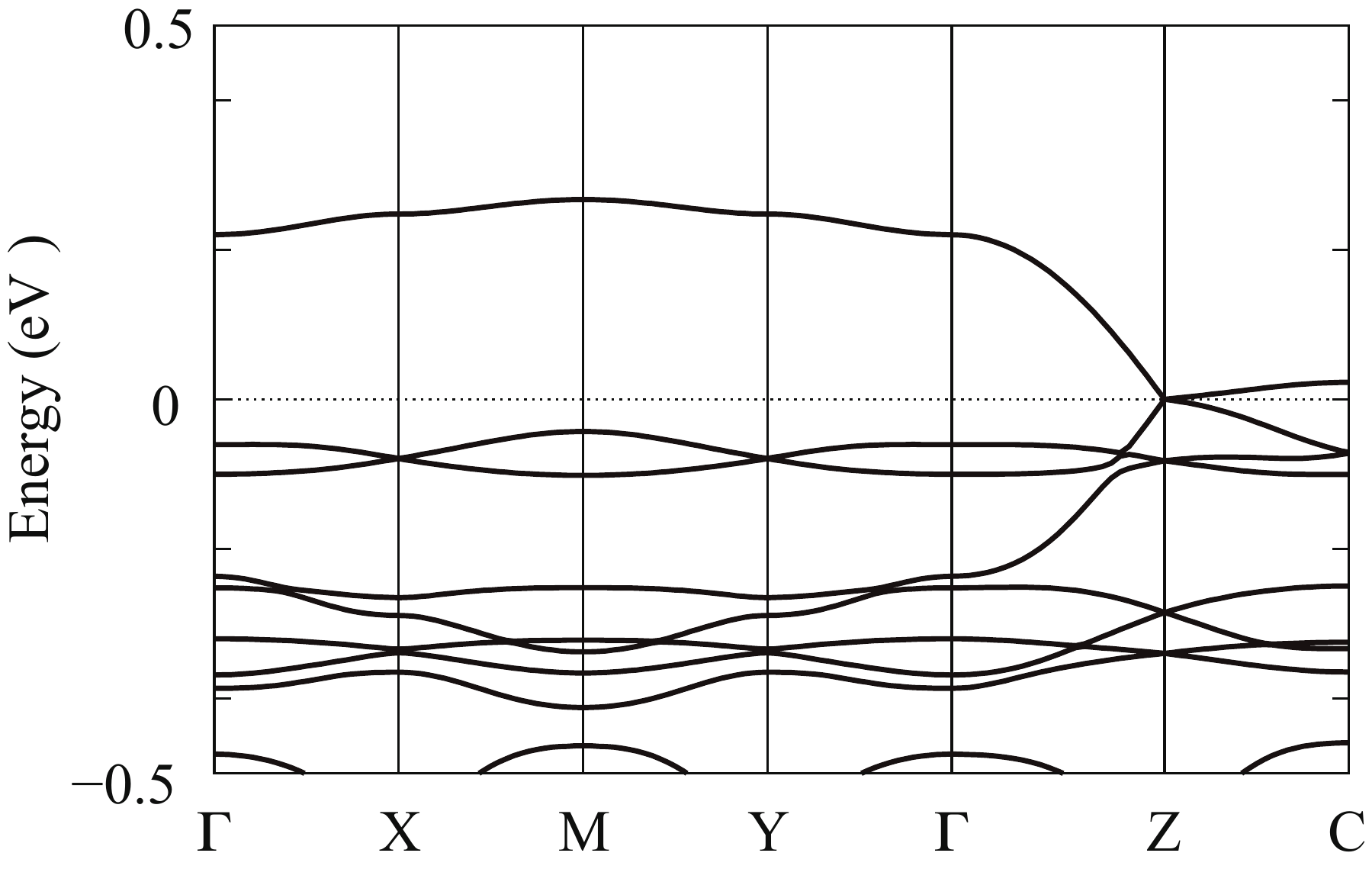}
 \caption{\label{Fig:band2} Band structure  of 
 (BEDT-TTF)Cu[N(CN)$_2$]$_2$ obtained by first-principles DFT calculations.
 The Fermi level is set to zero energy at the Dirac point $Z$ (dotted line).
}
\end{center}
\end{figure}
We move on to the experimental results.
Figure 4 shows the Arrhenius plot of the dc resistivity with the electric 
field parallel to three directions ($a^*$, $b$, and $c$-axes). 
The most conductive direction is the $c$-axis (crystal long axis) with the resistivity of
approximately 20 $\Omega$cm at RT.
The resistivities along the $b$ and $a^*$-axes are 4 and 5 orders higher 
than that along the $c$-axis, respectively; $\rho_c \ll \rho_b \leq \rho_{a^*}$.
The high anisotropy is consistent with the quasi-1D band dispersion.
The behavior of the resistivity in whole the temperature range measured 
below $\approx$ 200 K can be explained as an insulator with an excitation energy $E_a$
of approximately 0.13 eV (solid line in $E \parallel c$), which is comparable with the previous study\cite{ETDCA1}.
A slight non-linearity of the Arrhenius plot is observed above 200 K, of which
we confirmed high reproducibility by using several specimens.
This deviation above 200 K will be related to the non-ohmic behavior 
(current-dependent $\rho(T)$) observed recently\cite{ETDCA2}.

\begin{figure}[b]
\begin{center}
\includegraphics[width=6.5cm,pagebox=cropbox,clip]{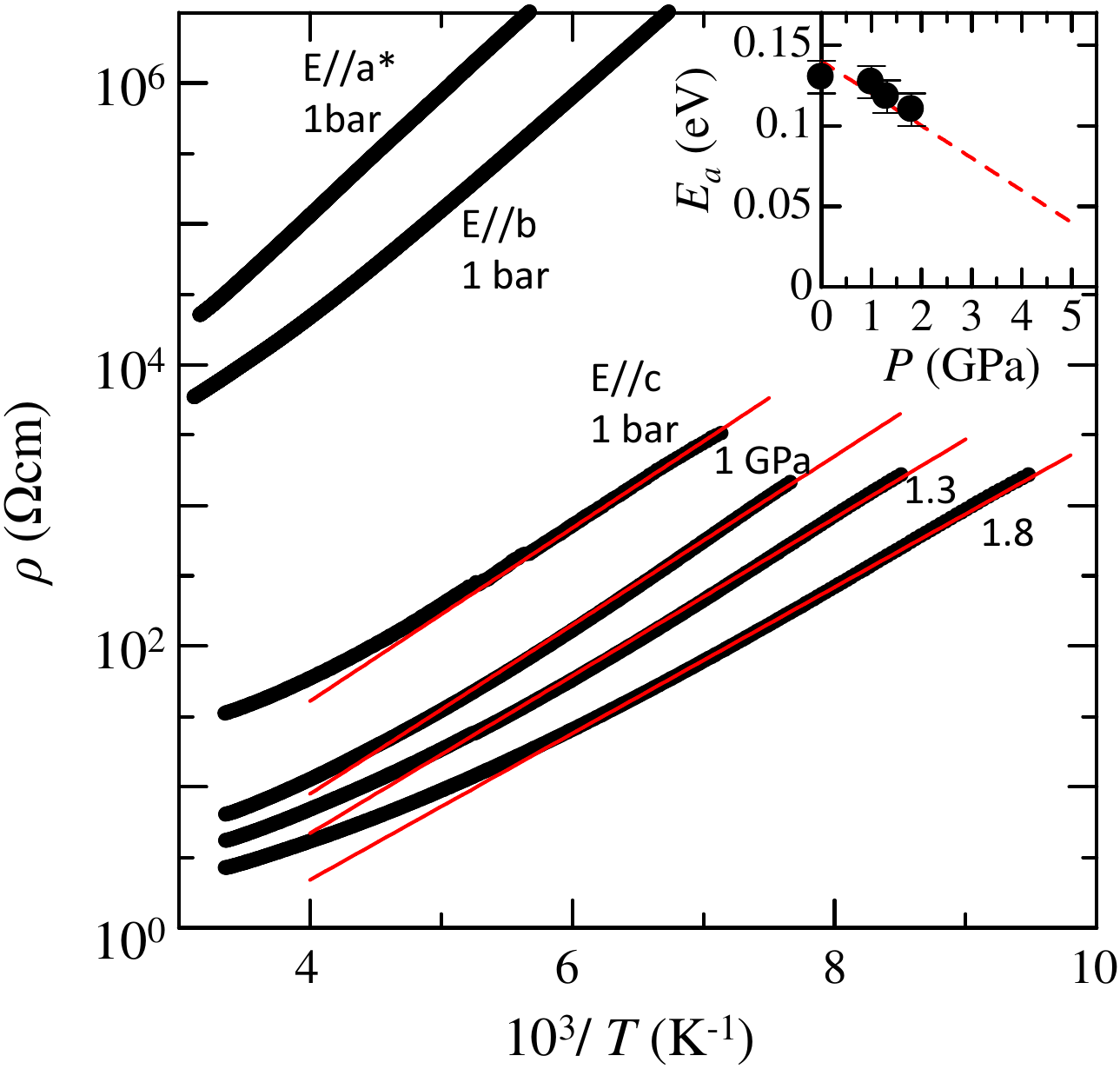}
 \caption{\label{Fig:Ap} (Color online) Arrhenius plot of the dc resistivity of (BEDT-TTF)Cu[N(CN)$_2$]$_2$.
The anisotropic resistivities along the three directions are measured in ambient pressure;
the resistivity under pressure is carried out in $E \parallel c$.
The (red) solid lines are obtained by means of a linear fitting below 
200 K, resulting in the pressure dependence of the activation energy (inset).
}
\end{center}
\end{figure}
The present salt has the 1/2-filled band, and thus a most reasonable interpretation of
the insulating behavior is the scenario as a Mott 
insulator by the strong on-site Coulomb interaction $U_\textrm{eff}$.
Since the localized site should be an ET molecule (monomer),
$U_\textrm{eff}=U_0-V$, where $U_0$ is the on-site Coulomb interaction for 
an ideally isolated ET and $V$ is the sum of the inter-site interactions.
$2E_a$ is equivalent to the Hubbard gap $E_g=U_\textrm{eff}-W$, where $W$ is the band width.
Applying $E_a=0.13$ eV obtained experimentally and $W=4|t-t'|=0.86$ eV estimated from the tight-binding model, we obtain $U_\textrm{eff}=2E_a+W=1.12$ eV.
The magnitude of $U_\textrm{eff}$ is almost comparable to those of other
monomer Mott insulators: 
0.78 eV in (ET)Ag$_4$(CN)$_5$\cite{Ag2} and 0.82 eV in $\zeta$-(ET)PF$_6$\cite{zeta}.

In the resistivity measured under pressure along the $c$-axis,
the insulating behavior is still observed up to 1.8 GPa.
As shown in the inset of Fig. 4, $E_a$ monotonically decreases with increasing pressure.
A rough extrapolation at the rate of $dE_a/dP \approx -20$ meV/GPa 
(broken line in the inset)
implies that pressure much higher than 5 GPa is needed to 
suppress the charge gap completely. 
The present salt seems to be robust against pressure,
compared with (ET)Ag$_4$(CN)$_5$, $dE_a/dP \approx -34$ meV/GPa\cite{AgCN2}.
Unfortunately, in the latter salt, inevitable disorder in the anion (C/N 
site occupancy) disturbs the realization of the DS state under pressure.
As a notable feature of the present salt, there is no such disorder in the anion Cu[N(CN)$_2$]$_2^-$.
Thus if the crystal structure is safely preserved under pressure, 
the node line DS state can arise in a weak limit condition of electron correlation.

The localized carrier of (ET)Cu[N(CN)$_2$]$_2$ has an $S=1/2$ spin degree 
of freedom as observed in ESR\cite{ETDCA1}. 
The strong 1D network of ET suggests the low-dimensional feature 
in the temperature dependence of the paramagnetic susceptibility ($\chi_\textrm{para}$).
Figure 5 shows $\chi_\textrm{para}(T)$ in 1 T (closed circles), 
which is obtained from $\chi_\textrm{dc}(T)$ (open circles) after 
subtracting a Curie term of about 1.1\% per ET molecule ($\chi_\textrm{imp}(T)$).
The amplitude of $\chi_\textrm{para}$, approximately 2.3$\times 10^{-4}$ emu/mol at 300 K, 
monotonically decreases with decreasing temperature.
Below $\sim$ 25 K, a steep exponential reduction of $\chi_\textrm{para}$ is observed,
indicating a spin-singlet ground state.
The inset of Fig. 5 shows $\chi_\textrm{para}(T)$ at low temperatures
with an Arrhenius-type fitting of $\exp(-\Delta/T)$ using $\Delta=90$ K.

Very recently a ferromagnetic transition at 13 K has been reported in 
wire-shape (ET)Cu[N(CN)$_2$]$_2$.\cite{ETDCA2}
However, reproducibility of the ferromagnetism has not been achieved as far as we 
examined it for several specimens with careful attention to avoiding contamination of by-products.
The magnetic ground state is an open question at present, but
our preliminary measurements of ESR at low temperatures show disappearance
of the paramagnetic signal below approximately 25 K, which is consistent 
with the present results of $\chi_\textrm{para}(T)$ and the spin singlet 
state at low temperatures. The ESR study will be reported elsewhere.

\begin{figure}
\begin{center}
\includegraphics[width=6.5cm,pagebox=cropbox,clip]{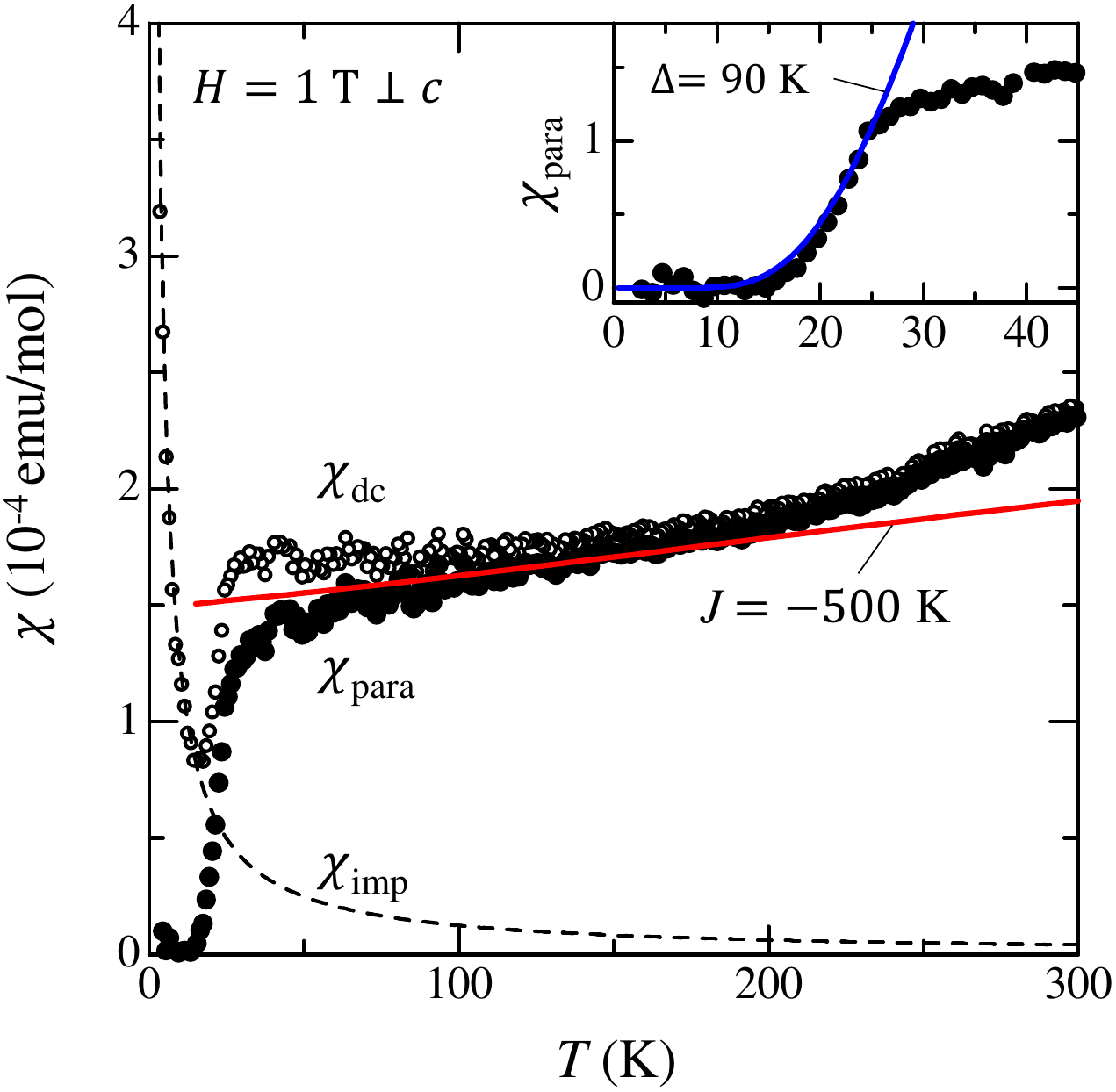}
 \caption{\label{Fig:kaiT} (Color online) The temperature dependence of the magnetic 
 susceptibility of (BEDT-TTF)Cu[N(CN)$_2$]$_2$. 
 $\chi_\textrm{para}$(closed circles) is obtained after subtracting $\chi_\textrm{imp}$ (broken curve)
 from  $\chi_\textrm{dc}$(open circles).
 Fitting for $\chi_\textrm{para}$ (red curve) is based on the 1D model\cite{BFfit}.
 Inset is the enlargement of $\chi_\textrm{para}(T)$ at low temperatures 
 with an Arrhenius fitting (blue curve).
}
\end{center}
\end{figure}

We finally discuss the paramagnetic behavior.
The gradual decrease of $\chi_\textrm{para}$ from 300 K down to 25 K is explained 
as a typical short-range ordering which appears in the lower 
temperature regime of a broad peak in the susceptibility of the low-dimensional 
localized spin systems. 
By ignoring the inter-chain interaction originating from $t'$,
the solid red curve in Fig. 5 is calculated on the basis of the $S=1/2$ 
Heisenberg antiferromagnetic spin model in the 1D lattice (Bonner-Fisher model\cite{BF,BFfit}, where the 
exchange interaction is defined as $-2J\mathbf{S}_i\cdot\mathbf{S}_j$)
with the intra-chain exchange interaction $J=-500$ K and the
Bohr magneton per ET of 1.0 $\mu_B$.
As another estimation of $J$, let the magnitude of $2J$ to be equal to 
the energy gap between the singlet and triplet states in the 1D Hubbard model,
$(-U_\textrm{eff}+\sqrt{U_\textrm{eff}^2+16t^2})/2$.
With the parameters $t=-0.19$ eV and $U_\textrm{eff}=1.12$ eV,
we obtain $|J|\approx 0.059$ eV($=650$ K), which is in good accordance with that
from the fitting to $\chi_\textrm{para}(T)$.
Although the fitting curve to $\chi_\textrm{para}(T)$ is not quite satisfactory,
this simple 1D model has suitable accuracy as a first-order approximation;
the fitting may be improved by taking the inter-chain exchange interaction ($t'$) into account.

In summary, we investigated the band structure, the resistivity, and the 
magnetic susceptibility in (ET)Cu[N(CN)$_2$]$_2$.
The uniform zigzag chain interaction results in the potential DS state with the nodal lines,
which could appear under pressure where the observed monomer Mott insulating state would be suppressed.
The paramagnetic insulating electronic properties in the 1/2-filled 
band is well explained in terms of the Mott insulator.
The sudden drop of $\chi_\textrm{para}$ indicates 
a spin-singlet (non-magnetic) ground state below 25 K.

\begin{acknowledgment}
The DFT calculations were conducted primarily at MASAMUNE at the Institute 
for Materials Research, Tohoku University, Japan.
This work was performed under the GIMRT Program of the Institute for 
Materials Research, Tohoku University (Proposal No. 202012-RDKGE-0034),
and partly supported by JSPS KAKENHI Grant No. 16K05747, 19K21860, 
19H01833, 21H05471, 22H01149, and 22H04459.
\end{acknowledgment}


\begin{thebibliography}{99}
\bibitem{Novoselov}K. S. Novoselov, A. K. Geim, S. V. Morozov, D. Jiang, 
M. I. Katsnelson, I. V. Grigorieva, S. V. Dubonos, and A. A. Firsov, Nature \textbf{438}, 197 (2005).
\bibitem{Katayama}S. Katayama, A. Kobayashi, and  Y. Suzumura, J. Phys. Soc. Jpn. \textbf{75}, 054705 (2006).
\bibitem{STF}T. Naito, R. Doi, and Y. Suzumura, J. Phys. Soc. Jpn. \textbf{89}, 023701 (2020).
\bibitem{BETS}S. Kitou, T. Tsumuraya, H.Sawahata, F. Ishii, K. Hiraki, T. 
Nakamura, N. Katayama, and H. Sawa, Phys. Rev. B \textbf{103}, 035135 (2021).
\bibitem{Pddddt2}R. Kato, H. Cui, T. Tsumuraya, T. Miyazaki, and Y. 
Suzumura, J. Am. Chem. Soc. \textbf{139}, 1770 (2017).
\bibitem{Ptdmdt2}B. Zhou, S. Ishibashi, T.Ishii, T. Sekine, R. Takehara, K. 
Miyagawa, K. Kanoda, E. Nishibori, and A. Kobayashi, Chem. Commun. 
\textbf{55}, 3327 (2019).
\bibitem{AgCN1}Y. Shimizu, A. Otsuka, M. Maesato, M. Tsuchiizu, A. Nakao, 
H. Yamochi, T. Hiramatsu, Y. Yoshida, and G. Saito, Phys. Rev. B 
\textbf{99}, 174417 (2019).
\bibitem{AgCN2}A. Kiswandhi, M. Maesato, S. Tomeno, Y. Yoshida,, Y. 
Shimizu, P. Shahi, J. Gouchi, Y. Uwatoko, G. Saito, and H. Kitagawa, 
Phys. Rev. B \textbf{101}, 245124 (2020). 
\bibitem{ETDCA1}H. H. Wang, U. Geiser, J. M. Williams, K. D. Carlson, A. M. 
Kini, J. M. Mason, J. T. Perry, H. A. Charlier, A. V. S. Crouch, J. E. 
Heindl, M. W. Lathrop, B. J. Love, D. M. Watkins, and G. A. Yaconi, Chem. 
Matt. \textbf{4}, 247 (1992).
\bibitem{ETDCA2}Y. Huang, T. Mitchell, D. C. Yost, Y. Hu, J. B. Benedict, 
J. C. Grossman, and S. Ren, Nano Lett. \textbf{21}, 9746 (2021).
\bibitem{BandCalc}T. Mori, A. Kobayashi, Y. Sasaki, H. Kobayashi, G. 
Saito, and H. Inokuchi, Bull. Chem. Soc. Jpn. \textbf{57}, 627 (1984).
\bibitem{PAW1994}P.~E. Bl\"{o}chl, Phys. Rev. B \textbf{50}, 17953 (1994).
\bibitem{DALCORSO_PP}A.~{Dal Corso}, Comp. Mater. Sci. \textbf{95}, 337 (2014).
\bibitem{QE2017}
P.~Giannozzi, O.~Andreussi, T.~Brumme, O.~Bunau, M.~B. Nardelli, M.~Calandra,
  R.~Car, C.~Cavazzoni, D.~Ceresoli, M.~Cococcioni, N.~Colonna, I.~Carnimeo,
  A.~D. Corso, S.~de~Gironcoli, P.~Delugas, R.~A. DiStasio, A.~Ferretti,
  A.~Floris, G.~Fratesi, G.~Fugallo, R.~Gebauer, U.~Gerstmann, F.~Giustino,
  T.~Gorni, J.~Jia, M.~Kawamura, H.-Y. Ko, A.~Kokalj,
  E.~K\"{u}{\c{c}}{\"u}kbenli, M.~Lazzeri, M.~Marsili, N.~Marzari, F.~Mauri,
  N.~L. Nguyen, H.-V. Nguyen, A.~O. de-la Roza, L.~Paulatto, S.~Ponc{\'{e}},
  D.~Rocca, R.~Sabatini, B.~Santra, M.~Schlipf, A.~P. Seitsonen, A.~Smogunov,
  I.~Timrov, T.~Thonhauser, P.~Umari, N.~Vast, X.~Wu, and S.~Baroni: J. Phys.
  Cond. Matter \textbf{29}, 465901 (2017).
\bibitem{GGA_PBE}J.~P. Perdew, K.~Burke, and M.~Ernzerhof, Phys. Rev. 
Lett. \textbf{77}, 3865 (1996).
\bibitem{q4terminal}A voltage terminal was also used as a current probe
because the corresponding current lead was disconnected during pressure application.
\bibitem{crystaldata}The crystallographic data at 296 K are:
formula C$_{14}$H$_8$S$_8$CuN$_6$, space group $C2/c$, $Z=4$, 
$a=16.756(3)$\AA, $b=13.975(3)$\AA, $c=10.340(2)$\AA, $\beta=124.145(10)$\textdegree, $V=2003.9(7)$\AA$^3$,
$d_\textrm{calc}=1.923$ g cm$^{-3}$. $R_1$(for $I>2\sigma(I)$, 1882 
reflections) = 0.041, $wR_2$(for all, 2306 reflections) = 0.1127, and
Goodness of fit (GOF) = 1.099.
The data at 100 K are:
$a=16.781(3)$\AA, $b=13.816(3)$\AA, $c=10.3114(16)$\AA, $\beta=123.816(5)$\textdegree, $V=1986.3(6)$\AA$^3$,
$d_\textrm{calc}=1.940$ g cm$^{-3}$. $R_1$(for $I>2\sigma(I)$, 2094 
reflections) = 0.0374, $wR_2$(for all, 2196 reflections) = 0.1184, and
GOF = 1.332.
\bibitem{comment_on_node} 
In the 1st BZ, always $\alpha_k \geq 0$. Thus, if $\beta_k \geq 0$,
the condition $E_{\pm}({\bm{k}})=0$ is fulfilled by the 
following two cases: (i) $\alpha_k=\beta_k=0$ or (ii) 
$t\alpha_k+t'\beta_k=0$ and $\cos(a'k_{a'}-b'k_{b'})-1=0$. We then 
obtain the two nodal lines as in the text. The node (i) linearly connects $Z$
with e.g., ($a'^*/2$, $-b'^*/2$, $c'^*/2$), while node (ii) is 
bending through e.g., ($-a'^*/2$, $-b'^*/2$, $c'^*/2$).
In case of $\beta_k <0$, it is numerically confirmed that there is no solution for $E_{\pm}({\bm{k}})=0$.
\bibitem{Ag2}A. Otsuka, Y. Shimizu, G. Saito, M. Maesato, A. Kiswandhi, 
T. Hiramatsu, Y. Yoshida, H. Yamochi, M. Tsuchiizu, Y. Nakamura, H. 
Kishida, and H. Ito, Bull. Chem. Soc. Jpn. \textbf{93}, 260 (2020).
\bibitem{zeta}H. -L. Liu, L. -K. Chou, K. A. Abboud, B. H. Ward, G. E. 
Fanucci, G. E. Granroth, E. Canadell, M. W. Meisel, D. R. Talham, and D. 
B. Tanner, Chem. Mater. \textbf{9}, 1865 (1997).
\bibitem{BF}J. C. Bonner and M. E. Fisher, Phys. Rev. \textbf{135}, A640 (1964).
\bibitem{BFfit}W. E. Estes, D. P. Gavel, W. E. Hatfield, and D. J. 
Hodgson, Inorg. Chem. \textbf{17}, 1415 (1978).

\end{thebibliography}
\end{document}